# Approximation of lateral distribution of atmospheric Cherenkov light at different observation levels for different primary particles. Applications for cosmic ray studies*

A. L. Mishev, S. Cht. Mavrodiev and J. N. Stamenov

Institute for Nuclear Research and Nuclear Energy, Bulgarian Academy of Sciences, 72, Tsarigradsko chaussee, Sofia 1284, Bulgaria

**Abstract:** This work summarizes the results presented at 29th International Cosmic Ray Conference in Pune India. Generally the aim of this work is to obtain the lateral distribution of the atmospheric Cherenkov light in extensive air showers produced by different primary particles in wide energy range and at several observation levels and to fit the obtained lateral distributions. Using one large detector and partially modified CORSIKA code version are obtained the lateral distributions of Cherenkov light flux densities at several observation levels for different particle primaries precisely at 536 g/cm$^2$ Chacaltaya, 700 g/cm$^2$ Moussala and 875 g/cm$^2$ Kartalska field observation levels for hadronic primaries and gamma quanta in the energy range $10^{11}$ eV-$10^{16}$ eV. On the basis of the solution of over-determined inverse problem the approximation of these distributions is obtained. The same model function for all the primaries is used and for the different observation levels. The different model parameters for the different primaries and levels are obtained. The approximations are compared with polynomial approximation obtained with different method. Both approximations are used for detector efficiency estimation for the different experiments in preparation and estimation of the accuracy of the reconstruction techniques.

At the same time inclined showers up to 30 degrees zenith angle are studied at Chacaltaya observation level. The obtained lateral distributions of vertical showers are compared with vertical showers model and the previously obtained approximation. This permits to adjust the reconstruction strategy and to study the model parameters behavior.

## 1. Introduction

In the field of high and ultra high energy astroparticle physics one can find several interesting problems to investigate. In one hand it is important to estimate with big precision the energy spectrum of the primary cosmic ray and to estimate their mass composition. This is important in attempt to build appropriate model for their origin and acceleration mechanisms. Above $10^{14}$ eV the only possibility for cosmic ray measurement is ground based i.e. the detection of the secondary cosmic ray. At the same time it is important to cover the gap between the ground based and the space-born gamma ray astronomy. Currently gamma-ray energies between 20 and 250 GeV are not accessible to space-borne detectors and ground-based air Cherenkov detectors. The scientific potential of the ground based gamma ray astronomy is enormous and covers both astrophysics and fundamental physics. In one hand it is possible to study objects such as supernova remnants, active galactic nuclei and pulsars. On the other hand the observations especially in the range of low energies will help to understand well the various acceleration mechanisms assumed to be at the origin of very high energy gamma quanta. One of the most convenient techniques in cosmic ray investigation is the atmospheric Cherenkov technique [1]. Several in preparation experiments such as TACTIC [2] or MAGIC [3] used the image technique i.e. the reconstruction of the Cherenkov image of the shower. The registration of the atmospheric Cherenkov light in extensive air showers (EAS) can be applied for both of the mentioned above problems. On the other hand it is possible to measure the Cherenkov light flux densities using several detectors. The reconstruction of lateral distribution of atmospheric Cherenkov light is on the basis of previously proposed method for mass composition and energy estimation of the primary particle [4, 5] and the solution of inverse problem using the REGN code [6]. Moreover this method permits to reject gamma induced showers from hadronic induced events. The method is proposed for Chacaltaya observation level of $536g/cm^2$. The method permits not only to reject gamma induced showers from hadronic induced events, but to estimate the mass composition of primary cosmic ray in the energy region around the "knee". This method is proposed for wave front sampling experiments using the atmospheric Cherenkov technique i.e. permitting the reconstruction of the lateral distribution of the atmospheric Cherenkov light in EAS. Such type of experiment is the HECRE proposal [7]. It's represent uniform set of detectors such as AIROBICC [8] (see fig.1). The aim of the experiment is the registration and reconstruction of the lateral distribution of the atmospheric Cherenkov light in EAS and afterwards on the basis of several criteria the estimation of the type and the energy of the initial primary particle. The HECRE experiment [7] is situated at high mountain altitude of 536 $g/cm^2$ at Chacaltaya cosmic ray station and as consequence it is near to the shower maximum. In fact this permit to reduce as much as possible the fluctuations in shower development and as result to obtain flatter characteristics .At the same time the registration of inclined showers gives several additional possibilities. Comparing vertical and inclined showers it is possible to estimate the relative fluctuations of the number of Cherenkov photons in EAS. Moreover the relative fluctuation of the total number of Cherenkov photons in the shower gives additional information about the mass composition. At the same is very important to provide registration of the EAS with quasi constant efficiency for the different incident particles. The distributions of interest are obtained with CORSIKA 6.003 code [9].

*This work is supported under NATO grant EAP.RIG. 981843 and FP6 project BEOBAL.

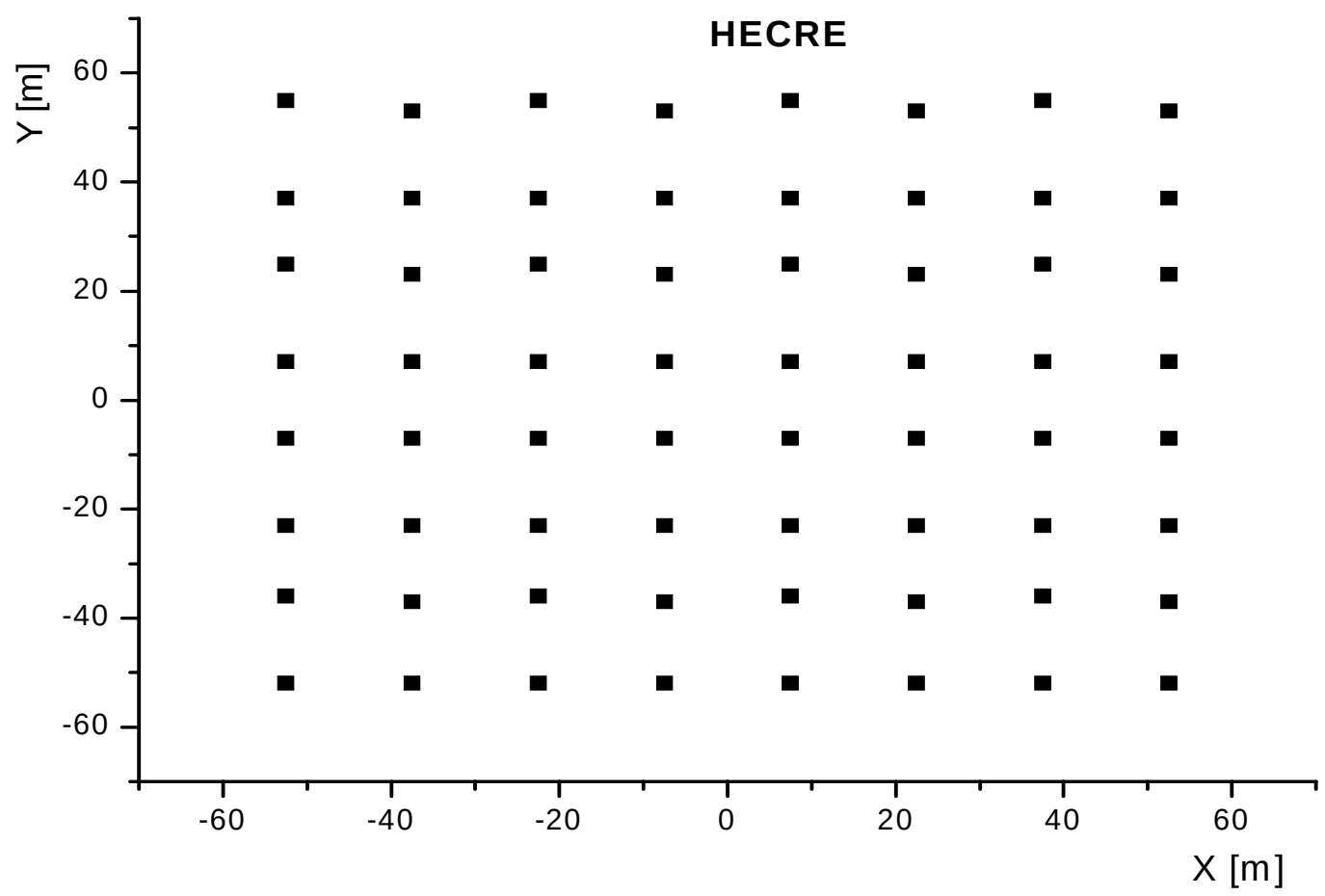

**Figure 1.** HECRE detector array

## 2. Results and Discussion

Using the CORSIKA 6.003 [9] code and QGSJET [10] and GHEISHA [11] as hadronic interaction models the lateral distribution of atmospheric Cherenkov light in EAS is obtained in wide energy range $10^{13}$-$10^{16}$ eV for different incident primaries. In attempt to reduce the statistical fluctuations we use one large detector, the aim to collect practically the totality of the Cherenkov photons in the shower. The simulation is carried out at high mountain altitude, actually observation level of 536 g/cm$^2$ at Chacaltaya observation level. In one hand it is possible to compare the obtained distributions with previously obtained vertical showers characteristics [12]. The simulated events are at least 500 per energy point and for different primaries such as proton, iron, oxygen, carbon and nitrogen. The bins change as a function of the distance from the shower axis. Near the shower up to 250 m from the axis the bin size is 10m because the densities are relatively high. After that the bins dimension are 20m and finally at larger distances they are 50m.

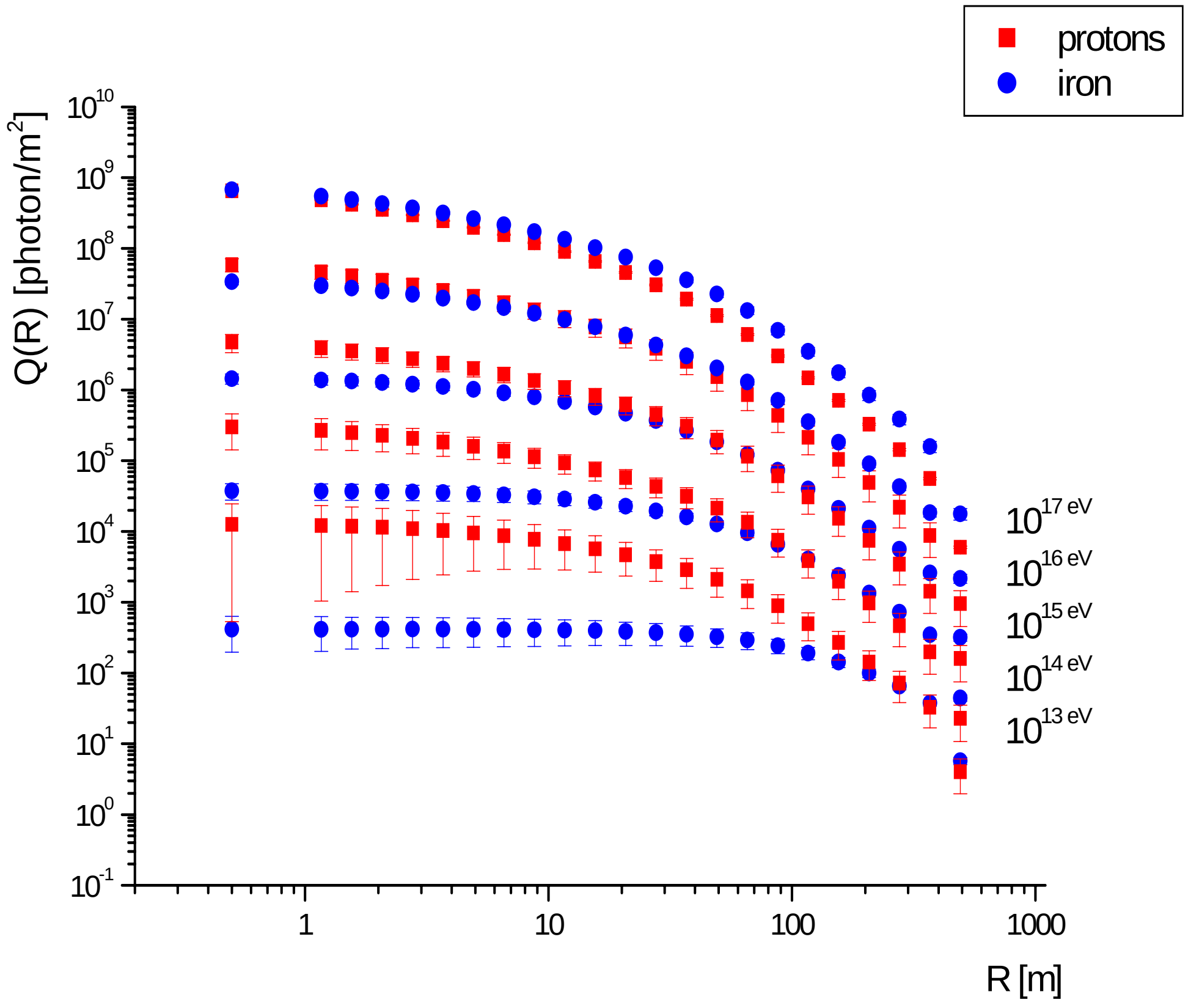

**Figure 2.** Difference between the lateral distributions of Cherenkov light flux in EAS initiated by primary protons and iron nuclei simulated with CORSIKA code in the energy range $10^{13}$-$10^{17}$ eV

The shape of the obtained distribution is similar but one observes several differences (fig.2). The main difference is that the iron produced lateral distribution is wider. Taking into account the previous experience this difference permits to obtain different model parameters in further approximation even in the case of the use of the same model function. Similar simulations are carried out for gamma quanta primaries. The results are shown in fig.3. In fact one can divide the energy range in two parts in attempt to use the obtained characteristics for solution of different problems. In the energy region around the “knee” and on the basis of previously proposed method [4, 5] using the REGN [6] code and inverse problem solution it is possible to use the obtained lateral distributions with the corresponding approximation for energy spectrum estimation and for mass composition estimation. In this case it is important to carry out the separation between the different hadronic primaries, thus it is necessary to obtain approximation with less as possible uncertainties in model parameters. The result for this approximation is presented in fig. 4.

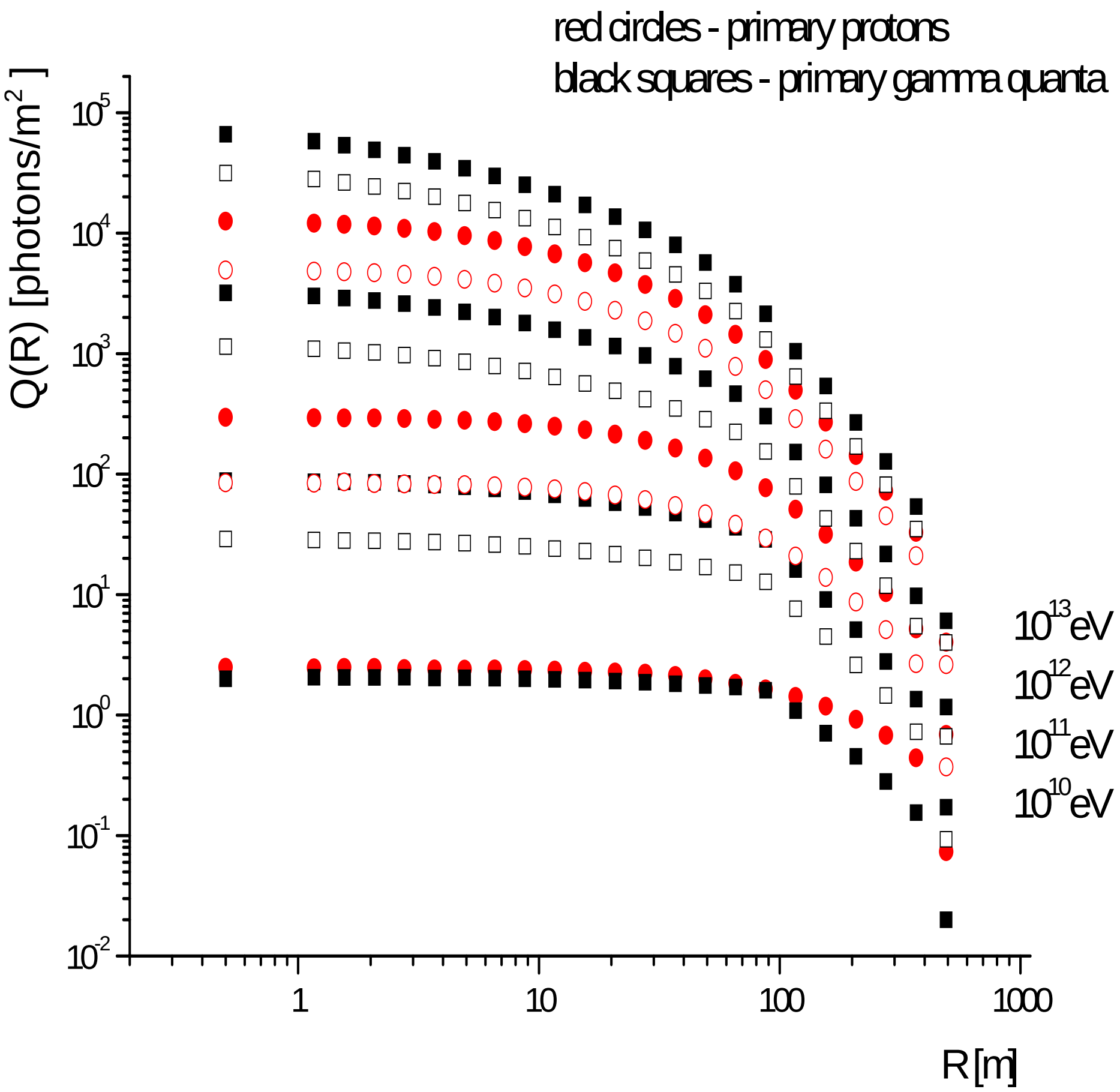

**Figure 3.** Lateral distributions of Cherenkov light flux in EAS produced by primary protons and gamma quanta in the energy range $10^{10}$ –$10^{13}$ eV

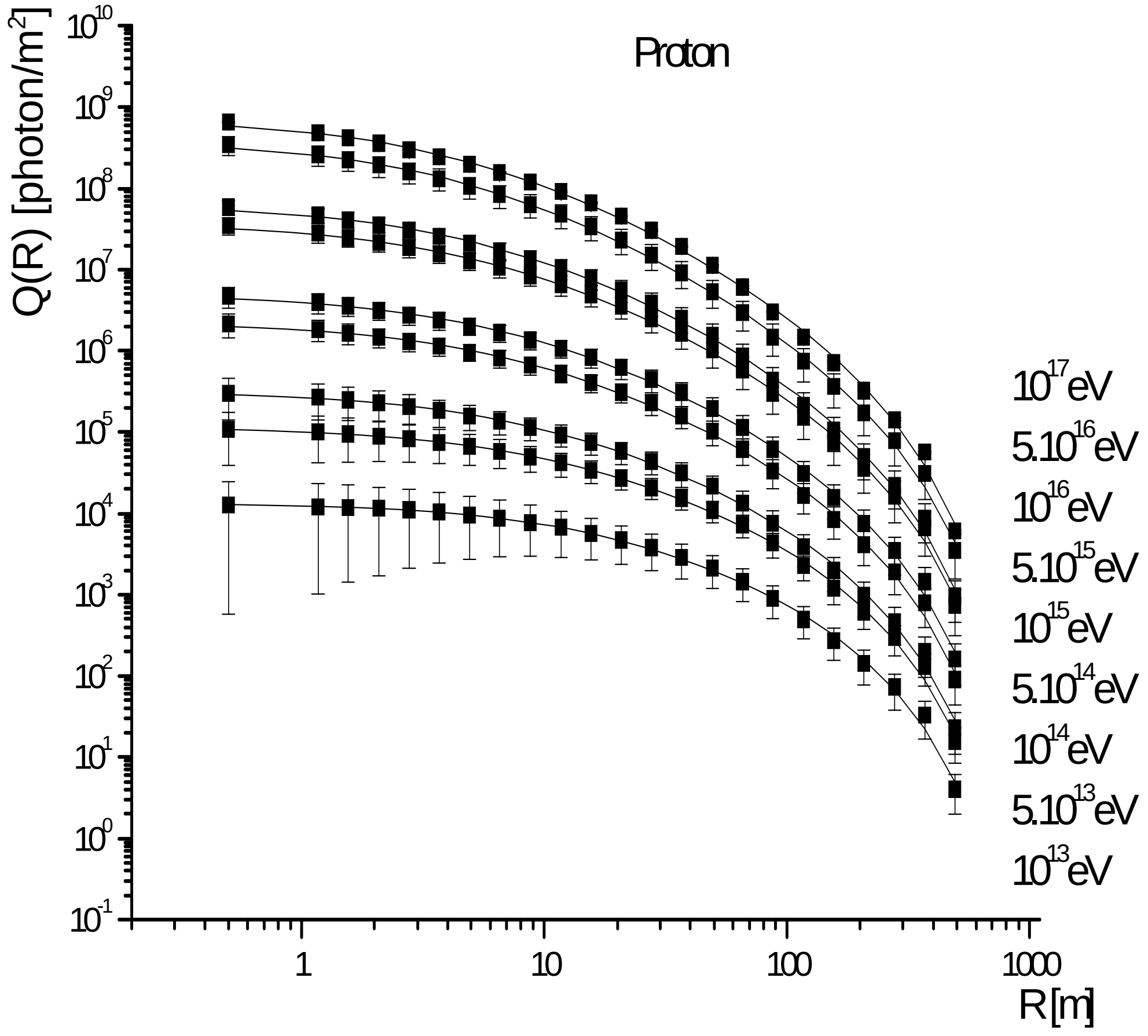

**Figure 4.** Lateral distribution of Cherenkov light initiated by primary protons in the energy range $10^{13}$ –$10^{17}$ eV simulated with the Corsika (scatter line) code and the obtained approximation (solid line) at 536 g/cm$^2$ observation level

The quality of the fit for iron and different hadronic primary induced showers is quite similar [4, 5, 13]. In fig. 5 is presented the lateral distribution of Cherenkov light flux in EAS initiated by primary iron nuclei with the corresponding approximation. The proposed approximation and the corresponding method is proposed for HECRE [6] experimental proposal at Chacaltaya observation level of 536 g/cm$^2$ in attempt to estimate the energy spectrum and mass composition around the "knee". Similar method is also proposed for ground based gamma ray astronomy [14, 15]. This method comparing to other experiments based on image Cherenkov technique use the wave front sampling telescope and reconstruction of lateral distribution of the registered Cherenkov light. Afterwards on the basis of the different $\chi^2$ it is possible to select only the incident gamma quanta, thus to reject the hadronic primaries. At the end we must point out that it is very important to obtain with big precision the end of the distribution. This permits in one hand to decrease the $\chi^2$ of the approximation and thus to obtain better results for separation of the different primaries. On the other hand gives the possibility to estimate better the energy of the primary particle, which reflects on model parameters uncertainties actually to reduce them and therefore on separation between the different primaries.

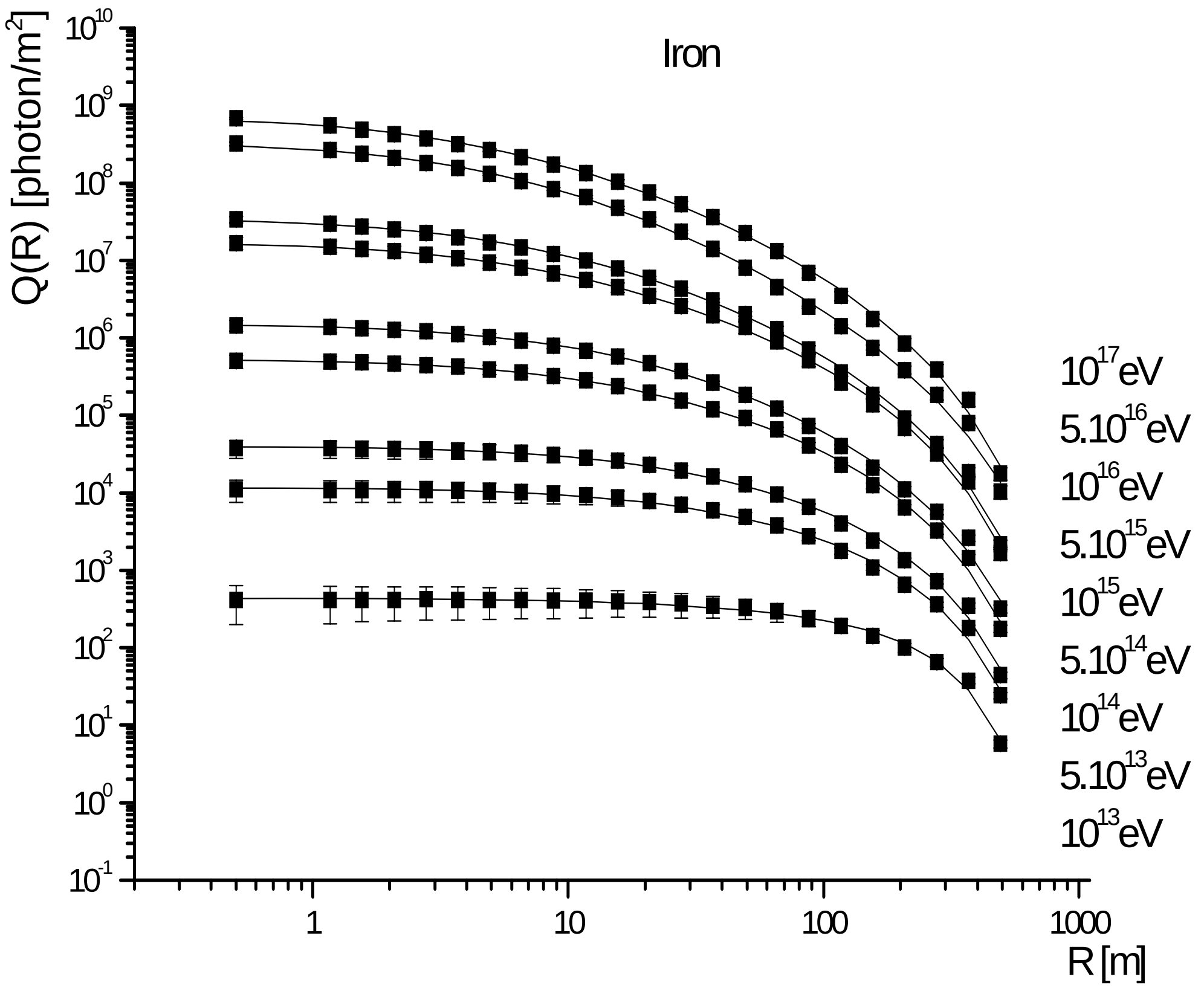

**Figure5.** Lateral distribution of Cherenkov light flux in EAS initiated by primary iron nuclei in the energy range $10^{13}$ –$10^{17}$ eV simulated with the Corsika (scatter line) code and the obtained approximation (solid line)

Another yet operational experiment based on atmospheric Cherenkov technique is the Ice Lake [16]. In this experiment one uses the reflected from snow surface Cherenkov light, which is measured by two spherical mirrors working in coincidence. For this experiment it is important to estimate precisely the energy threshold and the expected counting rate. This is important for partial calibration and the further data analysis. Therefore it is necessary to obtain the lateral distribution of atmospheric Cherenkov light at this observation level, actually 700 g/cm$^2$ and to proceed for detailed Monte Carlo simulation of the detector response for both, reflected by snow surface events and zenith measurements. This is the reason to carry out practically the same Monte Carlo simulations with Corsika 6.003 code [9] in the energy range around the “knee” with threshold energy of $10^{13}$ eV. In fig. 6 is shown the obtained with Corsika 6.003 code [9] with help of QGSJET [10] and GHEISHA [11] hadronic interaction models lateral distribution of atmospheric Cherenkov light in EAS produced by incoming proton primary particle with the corresponding approximation. At this observation level of 700 g/cm$^2$ the obtained approximation is used for fast Monte Carlo simulation of the detector response of Ice Lake experiment [16]. It is compared with the actual polynomial approximation, which was for the first time used for Plana experiment proposal detector response simulation [17]. The obtained result shows that the proposed model gives actually better as $\chi^2$ approximation, which permits to estimate with big precision the expected threshold of the experiment and counting rates. Moreover this study gives as additional result that the polynomial approximation gives some problems in a continuous energy spectrum because the relatively bad and difficult approximation of the polynomial fit parameters as a function of the energy. Contrarily the proposed model approximation [4] witch is with strong non linearity and taking into account the monotonic behavior of the model parameters as a function of the energy gives the possibility for more precise simulation of the detector response. In this analysis the data on Cherenkov light flux densities is used to simulate the amplitude spectrum and to estimate the energy of the air shower thus the primary particle energy. The obtained result of this study is not a topic of this work. Moreover the obtained approximation gives the possibility to simulate the telescope response for vertical events, which is important not only for partial calibration but for new method yet in development for atmospheric transparency estimation.

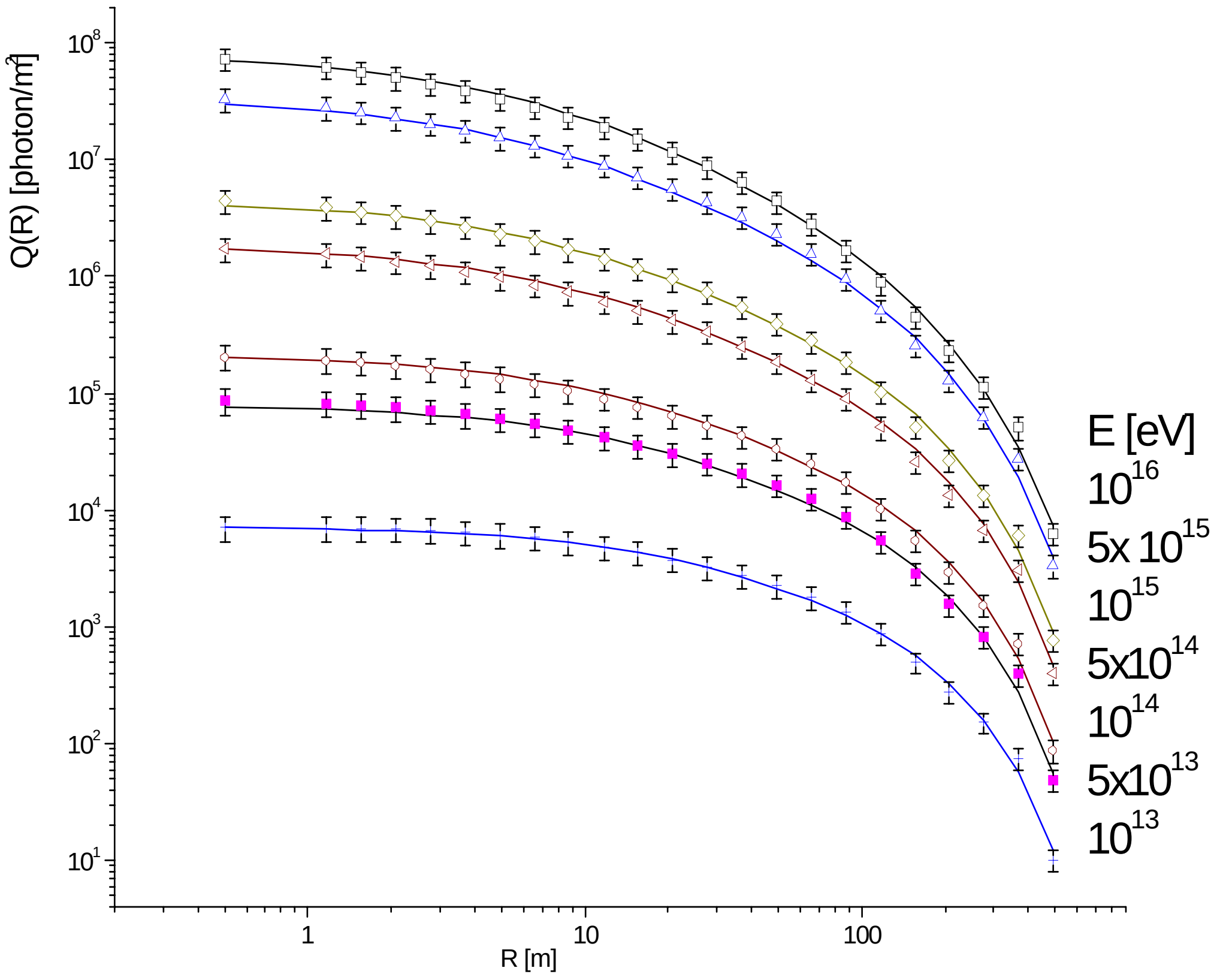

**Figure 6.** Lateral distribution of Cherenkov light initiated by primary proton in the energy range $10^{13}$ –$10^{16}$ eV simulated with the Corsika (scatter line) code and the obtained approximation (solid line) at 700 g/cm$^2$ observation level

Finally the same procedure is repeated for 875g/cm$^2$ observation level only for primary protons and gamma quanta. At 875 g/cm$^2$ observation the model calculations are carried out for the future Kartalska field experiment with design similar to the HOTOVO telescope [18]. Taking into account the less number of detectors for this experiment proposal and the actually applied number of detectors for the previously proposed model [4], the realistic topic for research is the ground based gamma ray astronomy. This is the reason to simulate the gamma quanta and protons induced showers presented in fig. 7. In this energy range and observation level the fluctuations is proton induced showers are higher and it is difficult to obtain flatter characteristics. The gamma quanta induced showers produces flatter lateral distribution of atmospheric Cherenkov light in EAS and as a consequence it is possible to approximate with big precision. Moreover the shape of the obtained lateral distribution is quite different. Even in this case we propose the same model function for approximation. Actually the obtained fit it is not with the same quality as for the hadronic primaries at other observation levels. The quality of the approximation of the lateral distribution of atmospheric Cherenkov light of proton induced showers is practically the same comparing to previous approximations. Even in this case taking into account the method, model parameters values and behavior as a function of the energy gives the possibility to reject the nuclei induced showers from electromagnetic showers. One possible improvement is to propose a different model function for the different primaries, which reflects on the facility of the data analysis. Summarizing, the proposed approximations and methodology with non linear fit and inverse problem solution gives excellent possibility for energy estimation of the primary cosmic ray with precision around 10-15% at different observation levels and to reject the hadronic induced showers from gamma quanta induced showers. Moreover using one universal model function gives additional advantages for reconstruction strategy and data analysis.

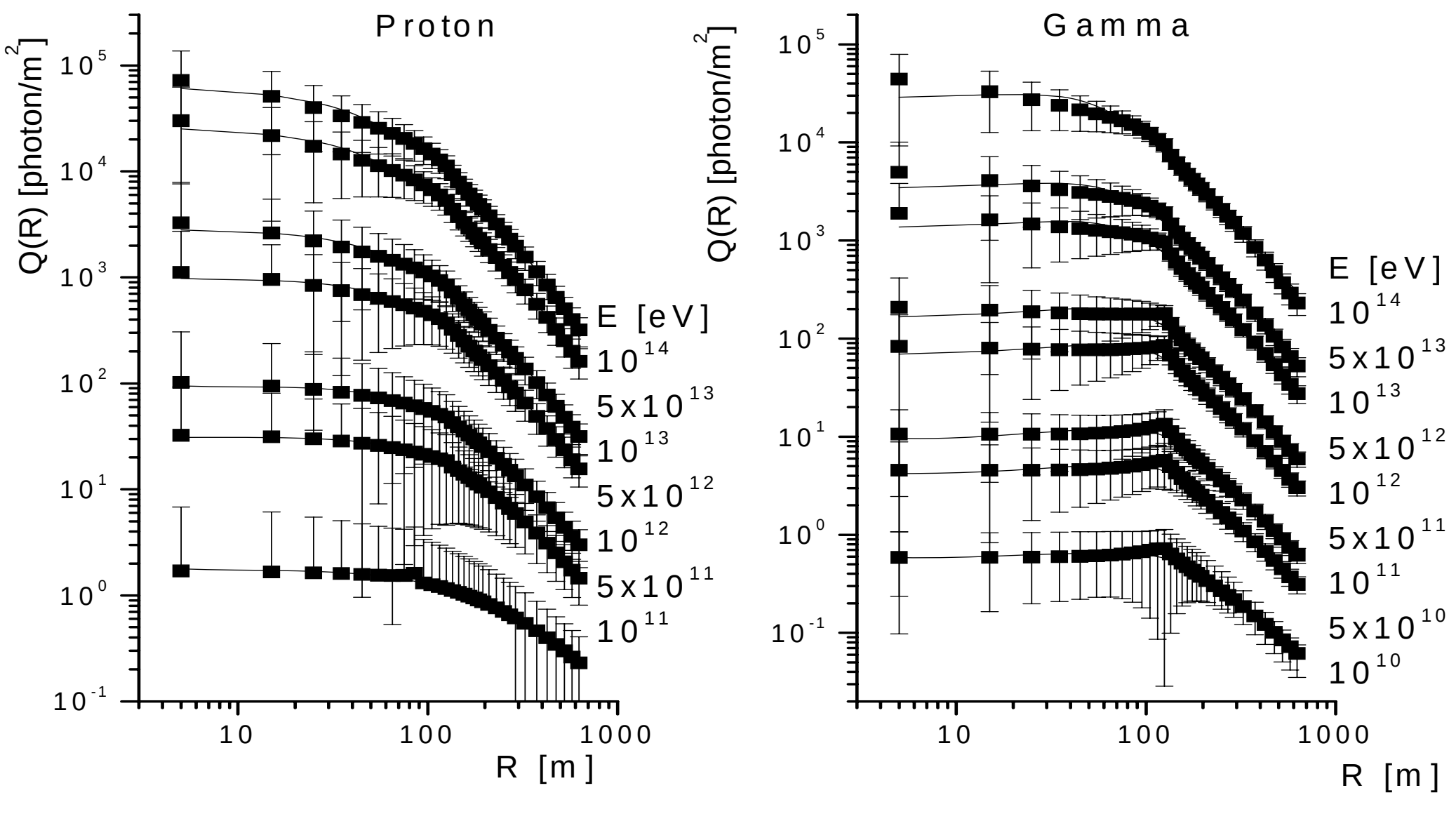

**Figure 7.** Lateral distribution of Cherenkov light initiated by primary proton and gamma quanta in the energy range $10^{11}$ –$10^{14}$ eV simulated with the Corsika (scatter line) code and the obtained approximation (solid line) at 875 g/cm$^2$

The characteristics on inclined showers are obtained as well. The registration of inclined showers with relatively good angular resolution gives several additional possibilities. Comparing vertical and inclined showers it is possible to estimate the relative fluctuations of the number of Cherenkov photons in EAS. At the same time it is possible to estimate the lateral distribution fluctuations as well. Moreover the relative fluctuation of the total number of Cherenkov photons in the shower gives additional information about the mass composition [19]. At the same is very important to provide registration of the EAS with quasi constant efficiency for the different incident particles [20, 21]. Using the CORSIKA 6.003 code and QGSJET [10] and GHEISHA [11] as hadronic interaction models the lateral distribution of atmospheric Cherenkov light is obtained in wide energy range $10^{13}$-$10^{16}$ eV for different primaries and for inclined showers. Once more in attempt to reduce the statistical fluctuations we use one large detector the aim to collect as much as possible of the Cherenkov photons in the shower. The simulation is carried out at high mountain altitude 536 g/cm$^2$ at Chacaltaya observation level. In one hand it is possible to compare the obtained distributions with previously obtained vertical showers [12]. On the other hand the obtained result gives the possibility to propose additional criteria for mass composition estimation and finally to adjust the methodology and model parameters for energy estimation and mass composition of primary cosmic ray [4, 5, 13]. In fig. 8 is presented the obtained lateral distribution of atmospheric Cherenkov light densities for proton incoming particle for vertical, 15 degrees and 30 degrees inclined showers and compared with vertical showers characteristics.

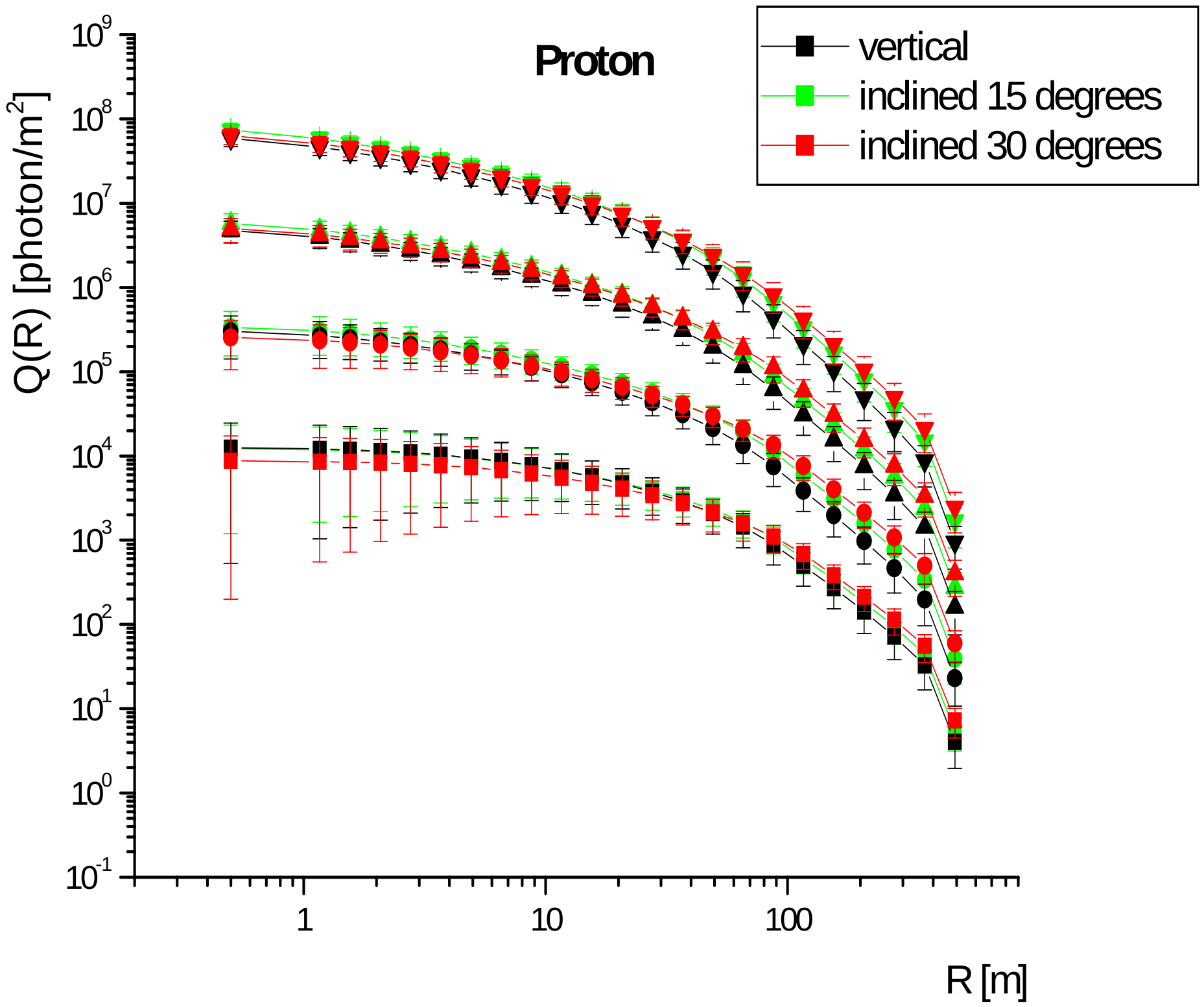

**Figure 8.** Lateral distribution of atmospheric Cherenkov light flux densities in the energy range $10^{13}$ eV – $10^{16}$ eV initiated by Proton induced inclined showers at Chacaltaya observation level

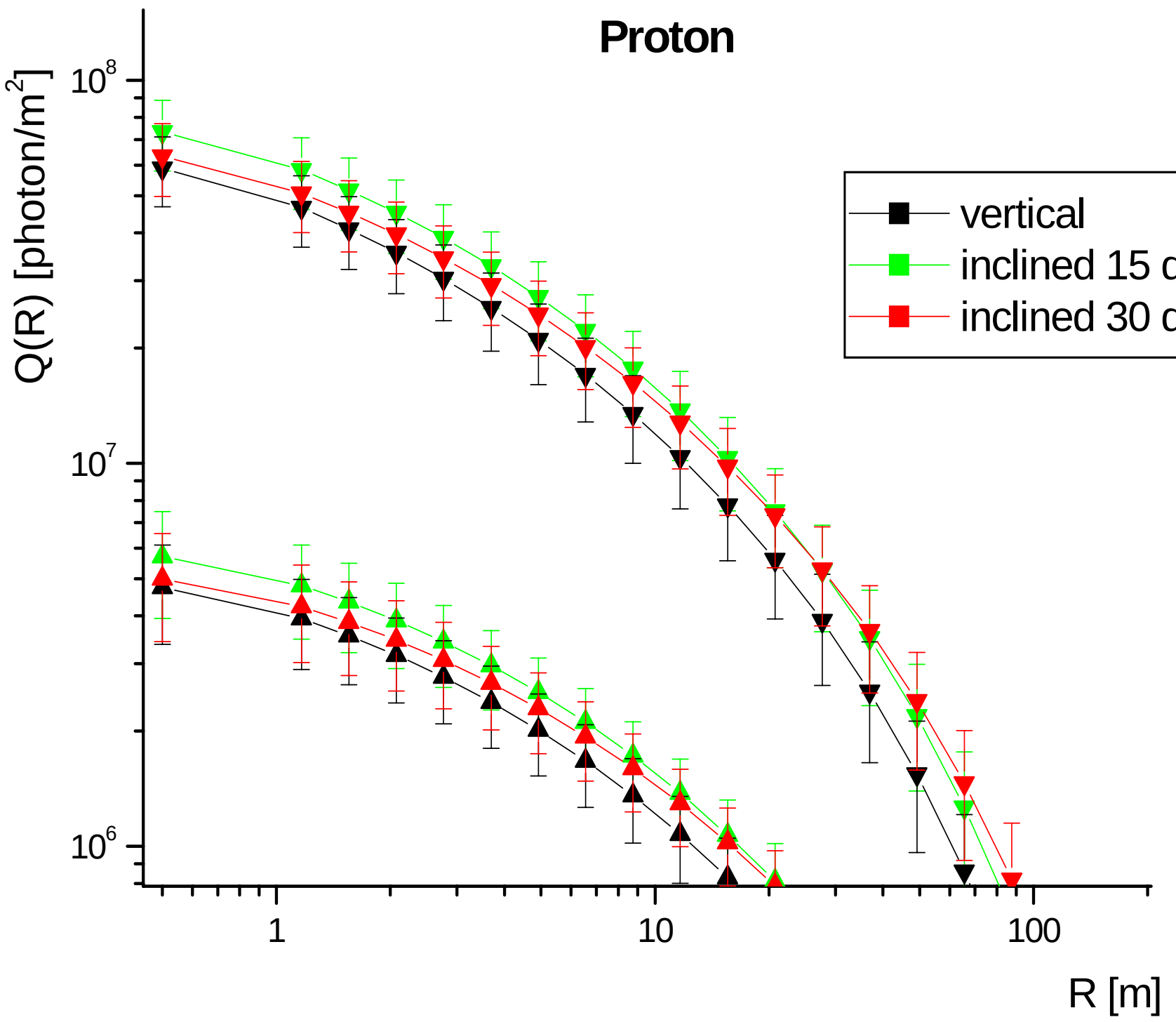

**Figure 9.** Lateral distribution of atmospheric Cherenkov light flux densities for $10^{13}$ eV and $10^{14}$ eV initiated by Proton induced inclined showers at Chacaltaya observation level

To present precisely the difference between the obtained lateral distributions in fig. 9 are shown the lateral distributions of atmospheric Cherenkov light flux densities for $10^{13}$ eV and $10^{14}$ eV initiated by Proton induced vertical, 15 and 30 degrees inclined showers. In fig. 10 is presented the obtained lateral distribution of atmospheric Cherenkov light densities for Helium induced showers for vertical, 15 degrees and 30 degrees inclined showers. In fig. 11 is presented the obtained lateral distribution of atmospheric Cherenkov light densities for Iron primary particle induced showers for vertical, 15 degrees and 30 degrees inclined showers. In this figure (fig. 11) is used logarithmic-linear scale towards to present the shape of the distribution.

It is easy to see that the lateral distributions of Cherenkov light initiated by showers inclined at different degrees are similar with small difference in the shape. The essential difference is observed in the values of the obtained densities and small difference in the observed fluctuations. As was expected as more the shower is inclined as more produces wider distribution (see fig. 8 and fig. 10). The observed systematic is the same for all simulated primaries. The obtained characteristics of inclined showers gives the possibility in one hand to build a strategy for mass composition estimation on the basis of registration of inclined with different incoming degrees and the relative fluctuations of the total number of Cherenkov photons in the shower at the given observation level. Moreover the obtained distribution gives the possibility to estimate the local densities of Cherenkov photons at given distance and so with the additional information of the zenith angle to select with constant efficiency the different primaries using the previously proposed selection parameter [20, 21].

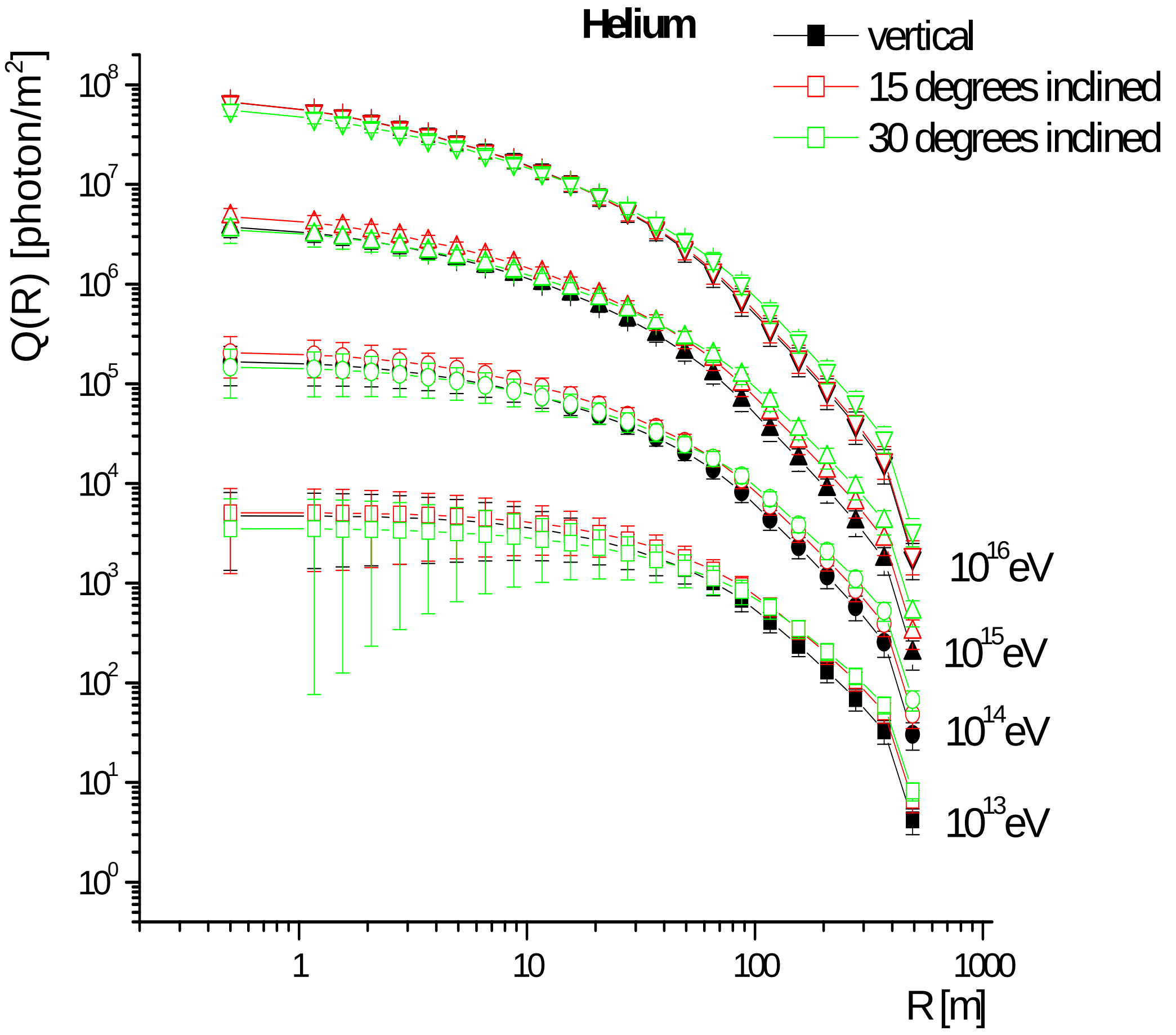

**Figure 10.** Lateral distribution of atmospheric Cherenkov light flux densities in the energy range $10^{13}$ eV – $10^{16}$ eV initiated by Helium induced inclined showers at Chacaltaya observation level.

In fig. 12 and fig. 13 is presented the difference between the lateral distributions of Cherenkov light flux in EAS initiated by primary proton and iron induced events for 15 degrees inclined showers in the energy range $10^{13}$ eV – $10^{16}$ eV. To present precisely this difference in fig. 13 are shown 15 degrees inclined proton and iron induced showers lateral distribution of Cherenkov light flux for $10^{13}$ eV.

At the same time the obtained lateral distribution is approximated with the proposed model function [4] for HECRE [7] experiment with similar results as was expected. This permits to adjust the model parameters, which is connected for uncertainties estimation. Approximating different primaries and inclined showers it is possible to obtain the limits of the model parameters their uncertainties. Moreover this is excellent check for the model behavior during the reconstruction procedure and inverse problem solution. Thus is very important to reduce the limits of the model parameters for the different primaries and as a result to estimate the mass composition of the primary cosmic ray with higher precision and efficiency.

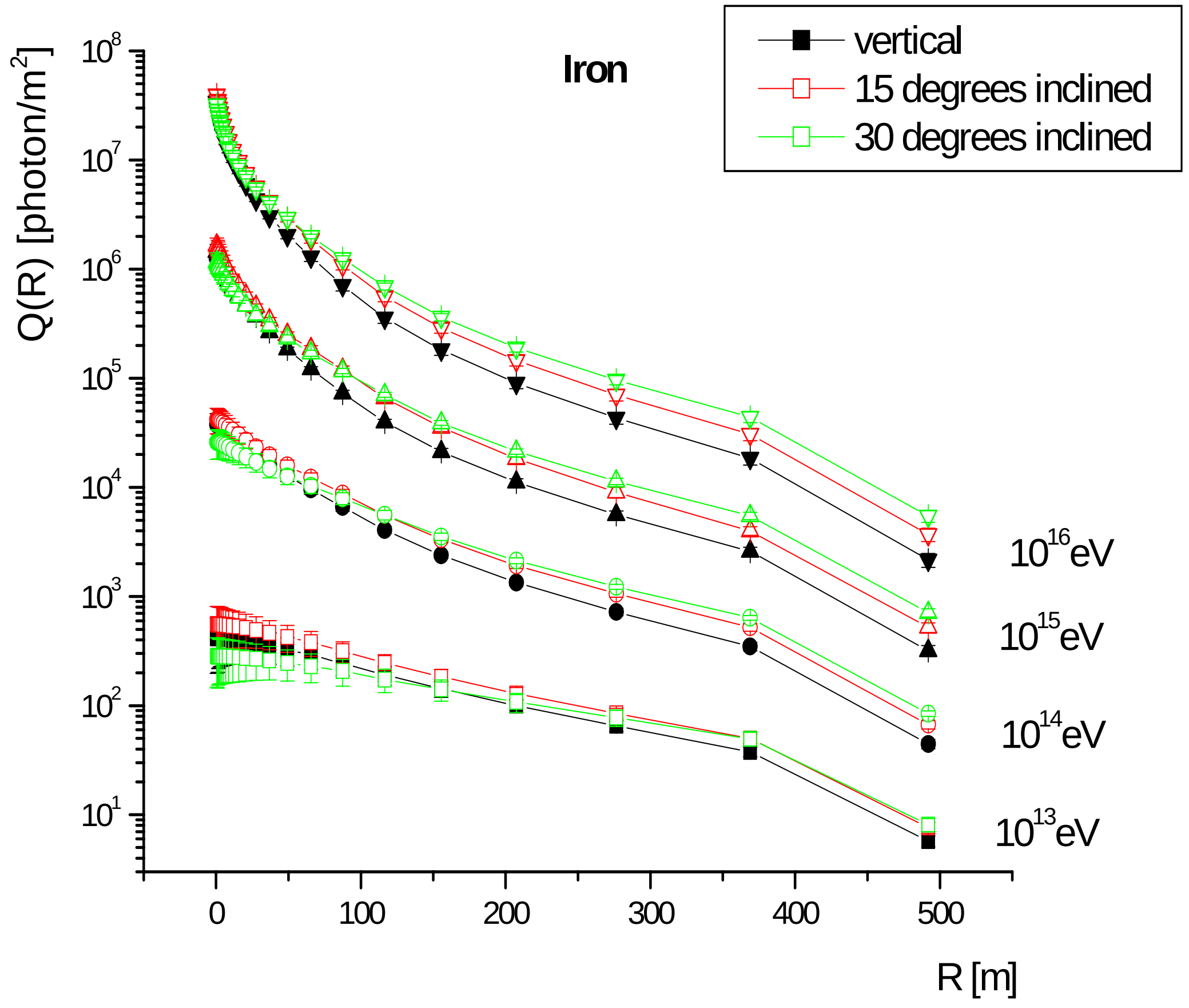

**Figure 11.** Lateral distribution of atmospheric Cherenkov light flux densities in the energy range $10^{13}$ eV – $10^{15}$ eV initiated by Iron induced inclined showers at Chacaltaya observation level.

And finally the obtained distributions are excellent check for the previously obtained methodology [4]. The good agreement between the model parameters and their behavior as a function of the energy of the initial particle show that the model and the proposed methodology are adequate and are usable for HECRE proposal. Moreover on the basis of the obtained results this method is usable for similar experiments using the atmospheric Cherenkov technique and wave front sampling of the Cherenkov light flux in EAS.
In the future it will be necessary to apply the same procedure for other primaries the aim being to obtain the similar results and looking towards reducing the model parameters values limits to investigate the mass composition of primary cosmic ray by groups. Moreover it is possible providing study of similar characteristics i.e. of different components such as electron, muon and hadronic component to make a multi-component analysis of the registered shower. This will permit more precise estimation of the reconstructed shower characteristics and thus to estimate the primary particle energy and mass.

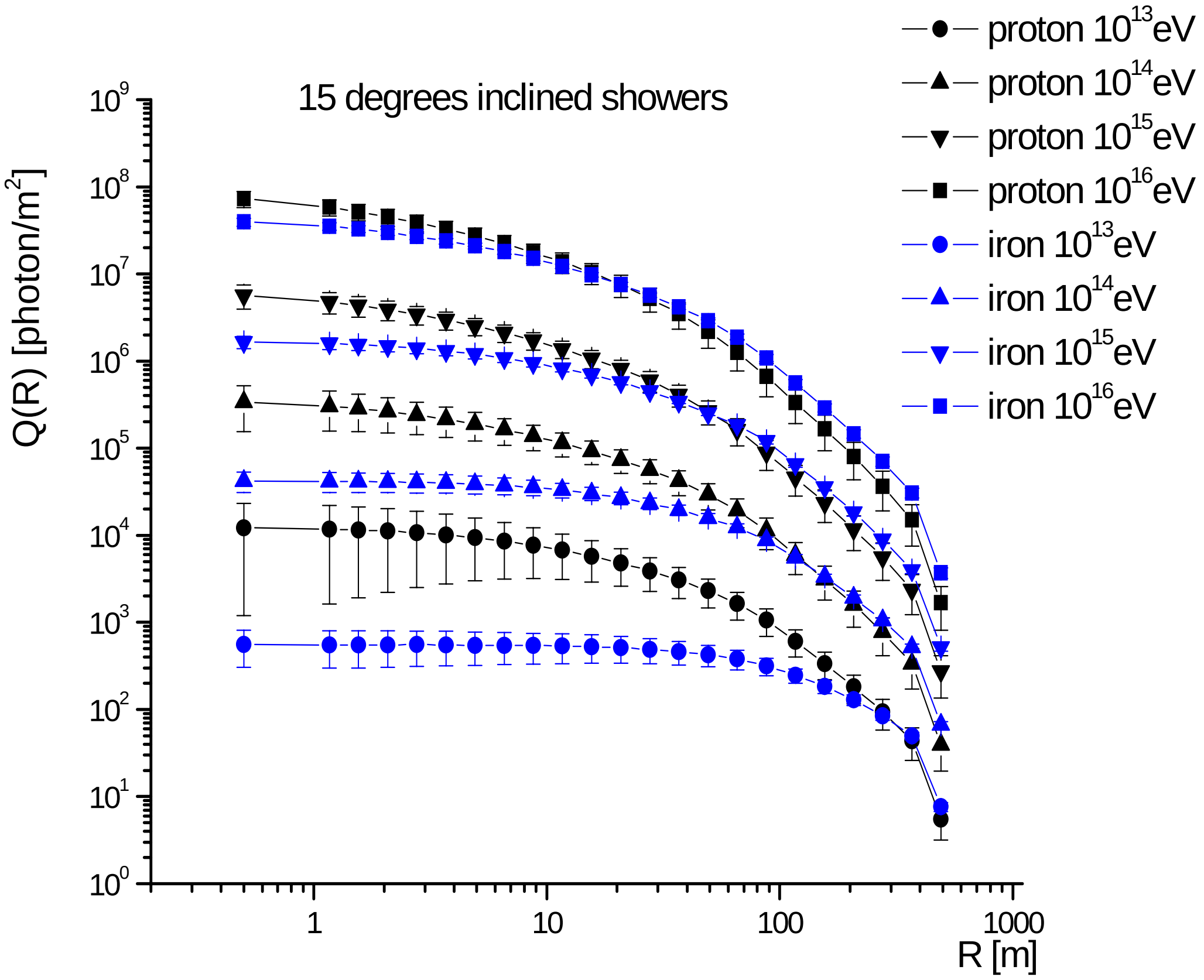

**Figure 12.** Difference between lateral distribution of atmospheric Cherenkov light flux densities in the energy range $10^{13}$ eV – $10^{15}$ eV initiated by Proton and Iron inclined showers at Chacaltaya observation level.

In fig. 13 the black line with circles corresponds to proton induced showers and the blue line corresponds to iron induced events. The same model function is used for approximation of inclined induced showers lateral distribution of Cherenkov light in EAS. The quality of the fit is the same and the behavior of the model parameters as function of the energy of the primary particle is quite similar. This permits to apply the obtained results for the discussed above problems.

Quite similar results are obtained for other zenith angles and for Helium and Carbon primaries. In the future an additional simulation for other primaries as Silicon and Oxygen is necessary towards to obtain more precise parameterization and to summarize the observed systematic. The final aim is to build a data bank of simulated data for the different primaries at this observation level, including inclined showers, which permits to propose a good reconstruction strategy.

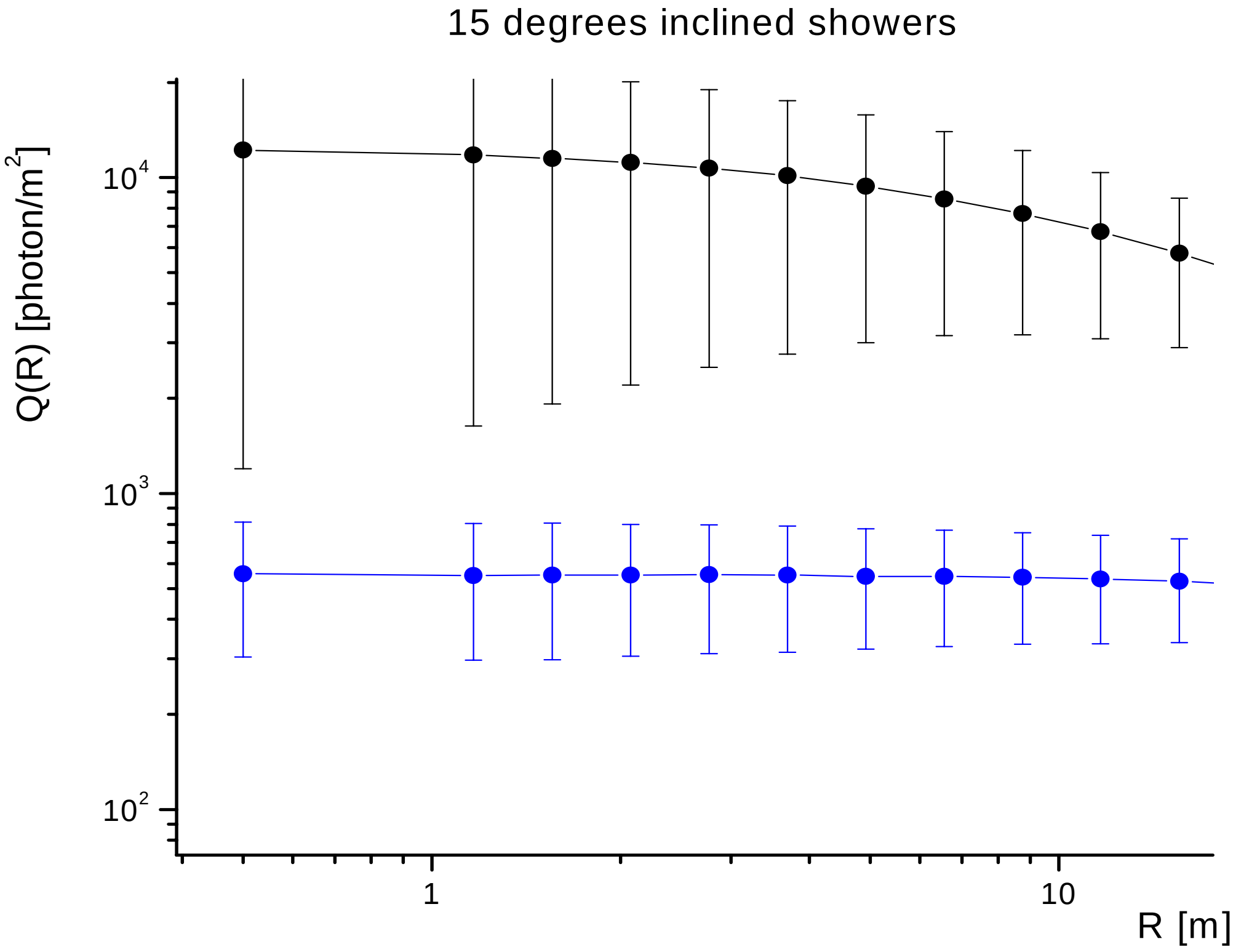

**Figure 13.** Difference between lateral distribution of atmospheric Cherenkov light flux densities for $10^{13}$ eV initiated by Proton and Iron inclined showers at Chacaltaya observation level.

**3. Conclusions**

Using simulated with Corsika 6.003 code data lateral distribution of atmospheric Cherenkov light flux densities in EAS are approximated with help of one model function in wide energy range and at different observation levels. The obtained approximation is applied for different primary particles precisely for primary proton, iron and Helium. Moreover the same model function was used for approximation of lateral distribution of Cherenkov light in EAS produced by primary gamma quanta at 875 g/cm$^2$ observation level which is important for the future Kartalska field experiment. This permits in one hand to build an appropriate model and method for shower characteristics reconstruction towards to estimate the energy and mass composition of the primary cosmic ray at different observation levels in wide energy range, experiment proposals etc… On the other hand one can use the obtained results for ground based gamma ray astronomy. At the same time using the same model function for the different observation levels and particles permits to build the same strategy for event reconstruction and to adjust the model using the behavior of model parameters as a function of the observation level and the energy of the primary particle. At the same time the lateral distribution of atmospheric Cherenkov light flux densities at Chacaltaya observation level are obtained with Corsika 6.003 code in the very interesting energy range around the "knee" for proton and iron primary particle and for inclined showers up to 30 degrees zenith angle. The obtained results are compared with previously obtained similar characteristic of vertical showers. They confirm the proposed model and methodology for event reconstruction at high mountain altitude based on atmospheric Cherenkov technique and permits to adjust several parameters and estimate the method constraints. Moreover the proposed approximation is the basis of the previously proposed model and permits full automatisation of event per event analysis.

**Acknowledgements**
We warmly thank to Dr. L. Alexandrov for the fruitful discussions during this study and Dr. Ivan Kirov for the suggestions. We also wish to thank the IT division of the INRNE for the assistance during the simulation and the given computational time of the computer cluster. The presentation of the results of this work was supported under NATO grant EAP.RIG. 981843 and FP6 project BEOBAL.

**Bibliography**

[1] A. Hillas Space Science Review, 75, 17 (1996)
[2] C. Bhat Proc. Perspectives in high energy astronomy and astrophysics (1996) Mumbai University Press
[3] A.Bariev et al., Max Plank Inst. report MPI-PhE/98-5 (1998)
[4] A. Mishev et al. Nucl.Instrum.Meth. A530, 359 (2004)
[5] S. Mavrodiev et al., 28th ICRC, Tsukuba (2003), 163
[6] L. Alexandrov, Journ. comp. math. and math. phys. 11, 36 (1971)
[7] O. Saavedra et al.,. Il Nuovo Cimento C vol. 24, 497 (2001)
[8] A. Karle et al., Astropart. Phys. 3, 321 , (1995)
[9] D. Heck et al., Report FZKA 6019 Forschungszentrum Karlsruhe (1998)
[10] N. Kalmykov et al., Phys. At. Nucl. 56 (3), 346 (1993)
[11] K. Werner, Phys. Rep. 232, 87, (1993)
[12] A. Mishev et al., 28th ICRC, Tsukuba (2003), 247
[13] A. Mishev et al., In Frontiers in Cosmic Ray Research **ISBN:** 1-59454-793-9, Nova Science 2005
[14] A. Mishev et al. Int. Journal of Modern Physics A, vol. 20 N 29, (2005) 7016-7019
[15] A. Mishev et al., ArXiv : astro-ph/0410118
[16] E. Malamova et al., 28th ICRC, Tsukuba (2003), 225
[17] I. Kirov et al., INRNE Ann. Rep. 1997-1998, 156 (1999)
[18] K. Berovsky et al., 24th ICRC, Calgary (1993), 1,
[19] M. Brankova et al, Il Nuovo Cimento C, Vol. 24, 525 (2001)
[20] M. Brankova et al., 27th ICRC, Hamburg (2001), 1968
[21] A. Mishev et al., 28th ICRC, Tsukuba (2003), 251